\documentclass[11pt,sort&compress]{elsarticle}

\journal{}
\usepackage{amsmath}
\usepackage{amssymb}
\usepackage{graphicx}
\usepackage{caption}
\usepackage{subcaption}
\usepackage[dvipsnames]{xcolor}
\usepackage[margin=1in]{geometry}
\usepackage{soul}
\usepackage{comment}
\usepackage{color}
\usepackage{xcolor}
\usepackage{etoolbox}
\usepackage{parskip}
\usepackage{microtype}
\usepackage{mathtools}
\usepackage{float}

\definecolor{lightblue}{rgb}{0.63, 0.74, 0.78}
\definecolor{seagreen}{rgb}{0.18, 0.42, 0.41}
\definecolor{orange}{rgb}{0.85, 0.55, 0.13}
\definecolor{silver}{rgb}{0.69, 0.67, 0.66}
\definecolor{rust}{rgb}{0.72, 0.26, 0.06}
\definecolor{purp}{RGB}{68, 14, 156}
\definecolor{joshua}{RGB}{251,220,127}

\colorlet{lightsilver}{silver!30!white}
\colorlet{darkorange}{orange!75!black}
\colorlet{darksilver}{silver!65!black}
\colorlet{darklightblue}{lightblue!70!black}
\colorlet{darkrust}{rust!85!black}
\colorlet{darkseagreen}{seagreen!85!black}
\definecolor{darksky}{HTML}{154c79}

\usepackage{bm}

\newcommand{\dd}{\text{d}}

\newcommand\Rey{\mbox{\textrm{Re}}}

\usepackage[colorlinks=true,linkcolor=darkrust,citecolor=darklightblue,urlcolor=darksilver]{hyperref}
\usepackage{cleveref}

\expandafter\def\csname ver@etex.sty\endcsname{}

\usepackage{autonum}

\usepackage[font=small,format=plain,labelfont=bf]{caption}

\crefname{appendix}{}{}

\begin{document}

\hypersetup{
  linkcolor=darkrust,
  citecolor=seagreen,
  urlcolor=darkrust,
  pdfauthor=author,
}

\begin{frontmatter}

\title{{\large\bfseries Targeted computation of nonlocal closure operators \\
via an
adjoint-based macroscopic forcing method}}

\author[1]{\vspace{-3ex}Jessie Liu}
\ead{jeliu@stanford.edu}
\author[2]{Florian Sch\"afer}
\author[2,3]{Spencer H.\ Bryngelson}
\author[4]{Tamer A.\ Zaki}
\author[1]{Ali Mani}

\address[1]{Department of Mechanical Engineering, Stanford University, Stanford, CA 94305, USA}
\address[2]{School of Computational Science \& Engineering, Georgia Institute of Technology, Atlanta, GA 30332, USA}
\address[3]{School of Aerospace Engineering, Georgia Institute of Technology, Atlanta, GA 30332, USA}
\address[4]{Department of Mechanical Engineering, Johns Hopkins University, Baltimore, MD 21218, USA}

\date{}

\begin{abstract}
Reynolds-averaged Navier--Stokes (RANS) closure must be sensitive to the flow physics, including nonlocality and anisotropy of the effective eddy viscosity. 
Recent approaches used forced direct numerical simulations to probe these effects, including the macroscopic forcing method (MFM) of Mani and Park (\href{https://doi.org/10.1103/PhysRevFluids.6.054607}{\textit{Phys. Rev.
Fluids} \textbf{6}, 054607 (2021)}) and the Green's function approach of Hamba (\href{https://doi.org/10.1063/1.2130749}{\textit{Phys. Fluids} \textbf{17}, 115102 (2005)}).
The resulting nonlocal and anisotropic eddy viscosities are exact and relate Reynolds stresses to mean velocity gradients at all locations. They can be used to inform RANS models of the sensitivity to the mean velocity gradient and the suitability of local and isotropic approximations.
However, these brute-force approaches are expensive. 
They force the mean velocity gradient at each point in the averaged space and measure the Reynolds stress response, requiring a separate simulation for each mean velocity gradient location. 
Thus, computing the eddy viscosity requires as many simulations as degrees of freedom in the averaged space, which can be cost-prohibitive for problems with many degrees of freedom. 
In this work, we develop an adjoint-based MFM to obtain the eddy viscosity at a given Reynolds stress location using a single simulation. 
This approach recovers the Reynolds stress dependence at a location of interest, such as a separation point or near a wall, on the mean velocity gradient at all locations. 
We demonstrate using adjoint MFM to compute the eddy viscosity for a specified wall-normal location in an incompressible turbulent channel flow using one simulation. In contrast, a brute-force approach for the same problem requires $N=144$ simulations (the number of grid points in the non-averaged coordinate direction). We show that a local approximation for the eddy viscosity would have been inappropriate. 

\end{abstract}

\end{frontmatter}

\section{Introduction}

Reynolds-averaged Navier--Stokes (RANS) models are widely used to simulate turbulent flows where direct numerical simulation (DNS) of the governing equations may be computationally cost-prohibitive. 
The flow variables are Reynolds decomposed into mean and fluctuating components, and the RANS equations govern the mean fields~\cite{pope2000turbulent}. 
However, an unclosed term involves the product of velocity fluctuations, commonly known as the Reynolds stress tensor. 
Further attempts to derive an exact evolution equation for the Reynolds stresses result in more unclosed terms, and hence the Reynolds stresses are typically modeled~\cite{spalart1992one,menter1994two,wilcox1998turbulence}. 

Recent works by \citet{hamba2005nonlocal} and \citet{park2021channel} have computed exact closure operators for Reynolds stresses. 
These closure operators can further be written in terms of generalized eddy viscosities that are nonlocal in space and time and anisotropic~\cite{hamba2005nonlocal}. 
The closure operators are exact in that the substitution of these operators back into the RANS equations results in exact mean quantities. 
Naturally the operators are problem-dependent, but they can be used to inform current RANS models of deficiencies in their eddy viscosity approximations and regions of sensitivity to the mean velocity gradient.

\citet{kraichnan1987eddy} derived an exact nonlocal and anisotropic expression for the Reynolds stress tensor using a Green's function. 
\citet{hamba2005nonlocal} modified the expression to be feasible for numerical implementation. 
\citet{hamba2005nonlocal} used the Green's function solution to a linearized formulation of the velocity fluctuation equation; the mean velocity gradient is treated as the source, and the velocity fluctuation is treated as the response. 
The generalized eddy viscosity is then formulated using Green's functions and velocity fluctuations. 
Because this approach needs the Green's function solution at each location in the averaged space, using a separate simulation for each location, computing the generalized eddy viscosity requires as many DNSs as degrees of freedom in the averaged space.
\citet{mani2021macroscopic} developed the macroscopic forcing method (MFM), a linear-algebra-based method for numerically obtaining closure operators. 
In MFM, one probes the closure operator by applying an appropriate forcing (not necessarily a Dirac delta function) to the governing equations and measures the averaged response. 
While MFM can obtain the exact generalized eddy viscosity similar to the approach of \citet{hamba2005nonlocal}, MFM can also obtain moments of the eddy viscosity using one simulation per desired moment. 
\citet{liu2023systematic} show how to use the limited information from a few low-order moments to model the eddy viscosity. 
The resulting eddy viscosity is nonlocal and matches the measured low-order moments, while the shape of its kernel approximately resembles the true kernel. 

For many applications, the exact eddy viscosity may be desired only within subregions of the domain where RANS models are particularly inaccurate, such as in regions of flow separation~\cite{jespersen2016overflow, probst2010comparison,park2022direct}. 
The generalized eddy viscosity at such locations can inform RANS models of the sensitivity of the Reynolds stresses at those locations to the mean velocity gradient at all locations. 
However, computing the generalized eddy viscosity using the aforementioned brute-force approaches requires forcing the mean velocity gradient at each location in the averaged space, entailing as many simulations as degrees of freedom in the averaged space. 

\begin{figure}[t]
    \centering
    \includegraphics[scale=1]{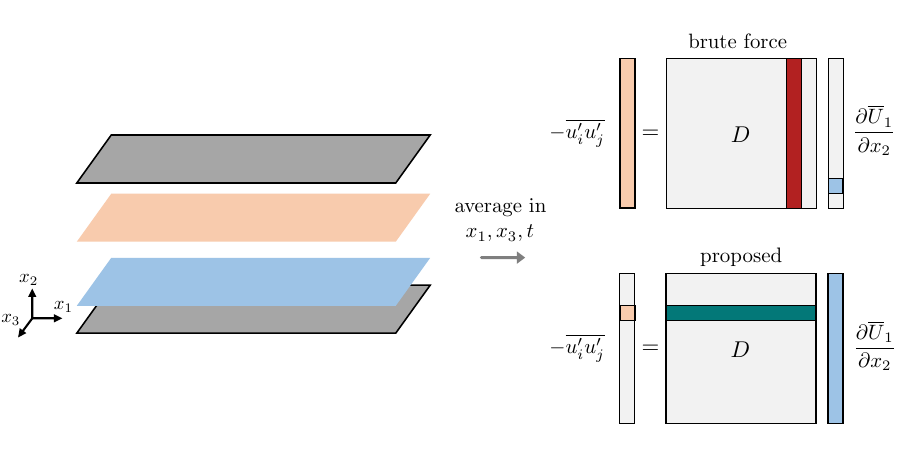}
    \caption{MFM illustration for a channel flow. 
    With a brute force approach, the mean velocity gradient, $\partial \overline{U}_1/\partial x_2$, is specified as an impulse at a specific location in $x_2$ (blue plane), which corresponds to activating one element of the mean velocity gradient vector as shown in the top right of the figure. A forced DNS is used to measure the Reynolds stress response, $-\overline{u_i'u_j'}$ at all locations. This recovers one column of the discretized eddy viscosity, $D$, and must be repeated for all mean velocity gradient locations.
    The proposed adjoint MFM obtains a more physically relevant row of $D$, relating the Reynolds stress at one location (orange plane) to the mean velocity gradient at all locations as shown in the bottom right of the figure.
    }
    \label{fig:overview} 
\end{figure}

We herein develop an adjoint-based method to compute the generalized eddy viscosity at a specific physical location using one simulation rather than via an expensive brute-force approach. 
\Cref{fig:overview} illustrates obtaining the generalized eddy viscosity for a canonical turbulent channel flow, using both the brute-force and our proposed approach. 
The mean velocity gradient is specified as an impulse at a specific location (\cref{fig:overview};~blue plane), and a forced DNS is used to examine the Reynolds stress response. 
One such brute-force simulation characterizes how the mean velocity gradient at a specific location influences the Reynolds stress at all locations, forming a column of the discretized eddy viscosity. 
The proposed adjoint-based approach characterizes how the Reynolds stress at a specific location (\cref{fig:overview};~orange plane) is influenced by the mean velocity gradient at all locations, forming a row of the discretized eddy viscosity and is more physically relevant. 
While we formulate the adjoint-based approach for Reynolds stress closures in this work, this approach can be used to inform closures more generally, including for scalar fluxes~\citep{hamba2004nonlocal}, compressible flows~\citep{lavacot2023nonlocality, wilcox1998turbulence}, and disperse multiphase flows~\citep{bryngelson2019quantitative}.

The adjoint-based formulation in this work can also aid in efficiently computing the eddy viscosity for the entire domain. 
Bryngelson and Sch\"afer et al.~\cite{bryngelson2023fast} reveal sparsity in the discretized eddy viscosity to establish Fast MFM, substantially reducing the number of simulations required to obtain the generalized eddy viscosity.
The adjoint-based method in this work enables a straightforward and computationally efficient way of recovering the operator rows (and columns) required for Fast MFM.
Through selective forcing, such that the output of each simulation contains information about multiple rows and columns of the discretized eddy viscosity, they developed a method to reconstruct the discretized eddy viscosity for the entire domain using substantially fewer simulations than a brute-force approach.

In \cref{problem formulation}, we define the generalized eddy viscosity and illustrate the cost of obtaining it using MFM. 
In \cref{adjoint MFM}, we develop adjoint MFM for obtaining the eddy viscosity for a specific Reynolds stress location. 
In \cref{numerical details}, we discuss the numerical details of the simulations. 
In \cref{results}, we compare MFM and adjoint MFM for obtaining the eddy viscosity at a specified wall-normal location in a turbulent channel flow at $\Rey_\tau = 180$ and show that a local approximation for the eddy viscosity is inappropriate.

\section{Problem formulation}\label{problem formulation}

Traditional RANS models~\cite{spalart1992one,menter1994two,wilcox1998turbulence} use the Boussinesq approximation~\cite{boussinesq1877essai} in which there are two underlying assumptions: 1) The length and time scales of the underlying velocity fluctuations are much smaller than that of the mean velocity fields, and hence the mixing by the turbulent fluctuations is assumed to be local; 
2) The mixing by the underlying fluctuations is assumed to be isotropic; hence, the Reynolds stress tensor and mean strain rate tensor are aligned. 
Under the Boussinesq approximation, an analogy is drawn to Brownian motion, for which random molecular mixing is modeled using a diffusive flux, and the Reynolds stress is modeled in terms of a scalar eddy viscosity and the mean velocity gradient. 
However, for turbulent flows, the underlying assumptions of the Boussinesq approximation are often invalid~\cite{corrsin1975limitations}.

\citet{hamba2005nonlocal} developed an exact closure for the Reynolds stress, $-\overline{u_i'u_j'}$, using a generalized eddy viscosity:
\begin{gather}
    \label{eq:generalized eddy viscosity}
    -\overline{u_i'u_j'}(\mathbf{x}, t) = \int_{\mathbf{y},\tau} D_{ijkl} (\mathbf{x}, \mathbf{y}, t, \tau) \frac{\partial \overline{U}_l}{\partial x_k} \bigg|_{\mathbf{y},\tau} \dd\mathbf{y} \dd\tau, 
\end{gather}
where $D_{ijkl}$ is a nonlocal and anisotropic eddy viscosity, and $\overline{U}_l$ is the mean velocity. 
The eddy viscosity is 1) spatiotemporally nonlocal in that the Reynolds stress depends on the mean velocity gradient at all points in space and time and 2) anisotropic in that the Reynolds stress tensor and velocity gradient tensor are not necessarily aligned. 

For the statistically stationary turbulent channel flow considered in this work, averaging is taken in time and over the homogeneous streamwise ($x_1$) and spanwise ($x_3$) directions. 
The simplified eddy viscosity is
\begin{gather}
    -\overline{u_i'u_j'}(x_2) = \int D_{ij21} (x_2, y_2) \frac{\partial \overline{U}_1}{\partial x_2} \bigg|_{y_2} \dd y_2,
\end{gather}
where $x_2$ is the wall-normal direction. 
\citet{hamba2005nonlocal} and \citet{park2021channel} computed the generalized eddy viscosity, $D_{ij21}(x_2,y_2)$, for a turbulent channel flow at $\Rey_\tau = 180$. 
\citet{hamba2005nonlocal} used the Green's function solution to a linearized equation for the velocity fluctuations. 
\citet{park2021channel} used inverse MFM (IMFM), where forcing is added to the governing equations to maintain a pre-specified mean velocity gradient. 
For computing the generalized eddy viscosity, \citet{liu2023systematic} showed that the two approaches are equivalent. 
However, \citet{hamba2005nonlocal} further performed averaging of $D_{ij21}$ to enforce symmetry in the Reynolds stress tensor, e.g., $(D_{2121} + D_{1221})/2$, whereas \citet{park2021channel} did not.
We discuss IMFM in this work, although one can also use the approach of \citet{hamba2005nonlocal}. 

\citet{park2021channel} simultaneously solve the incompressible Navier--Stokes equations:
\begin{subequations}
    \begin{align}
        \label{eq:NS momentum}
        \frac{\partial u_i}{\partial t} + \frac{\partial u_ju_i}{\partial x_j} &= -\frac{\partial p}{\partial x_i} + \frac{1}{\Rey}\frac{\partial^2 u_i}{\partial x_j \partial x_j} + r_i, \\
        \label{eq:NS mass}
        \frac{\partial u_i}{\partial x_i} &= 0,
    \end{align}
\end{subequations}
where $\Rey$ is the Reynolds number, $p$ is the fluctuating pressure, and $r_i$ is a body force, which for turbulent channel flow is the nondimensionalized mean pressure gradient, $r_i=(1,0,0)$, and the generalized momentum transport (GMT) equations:
\begin{subequations}
    \begin{align}
        \label{eq:GMT momentum}
        \frac{\partial v_i}{\partial t} + \frac{\partial u_jv_i}{\partial x_j} &= -\frac{\partial q}{\partial x_i} + \frac{1}{\Rey}\frac{\partial^2 v_i}{\partial x_j \partial x_j} + s_i, \\
        \label{eq:GMT mass}
        \frac{\partial v_i}{\partial x_i} &= 0,
    \end{align}
\end{subequations}
where $u_j$ is the advection velocity obtained from the Navier--Stokes equations, $v_i$ is a transported vector field, $q$ is a generalized pressure to ensure that $v_i$ is solenoidal, and $s_i$ is the IMFM forcing (and must satisfy $s_i = \overline{s}_i$). 
In this formulation, the eddy viscosity is
\begin{gather}
    -\overline{u_i'v_j'}(x_2) = \int D_{ij21} (x_2, y_2) \frac{\partial \overline{V}_1}{\partial x_2} \bigg|_{y_2} \dd y_2.
\end{gather}

\begin{figure}[t]
    \centering
        \begin{subfigure}[t]{0.5\textwidth}
        \centering
        \includegraphics[scale=1]{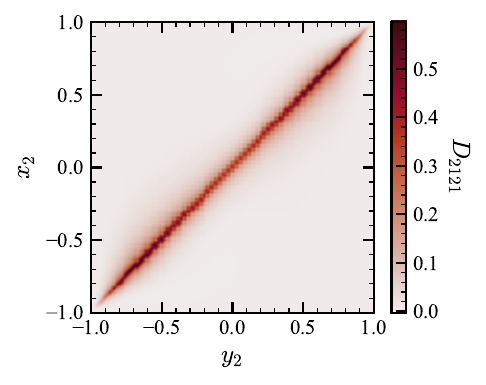}
        \caption{$D_{2121}(x_2,y_2)$}
        \label{fig:kernel_D2121}
     \end{subfigure}
     \begin{subfigure}[t]{0.49\textwidth}
         \centering
         \includegraphics[scale=1]{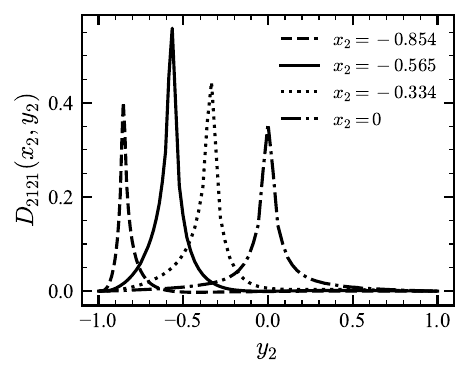}
         \caption{Rows of $D_{2121}$ for $x_2$ as labeled}
         \label{fig:D2121_slice}
     \end{subfigure}
     \caption{$D_{2121}(x_2,y_2)$ component of the generalized eddy viscosity for turbulent channel flow at $\Rey_\tau=180$. The rows of $D_{2121}$ represent the dependence of the shear component of the Reynolds stress, $-\overline{u_2'u_1'}(x_2)$, on $\partial \overline{U}_1/\partial x_2|_{y_2}$. (a) is reproduced from \citet{park2021channel} with author permission; (b) uses data from \citet{park2021channel}.
     }
\end{figure}

\Cref{fig:kernel_D2121} shows the $D_{2121}$ component of the eddy viscosity reproduced from \citet{park2021channel}.
This component represents the dependence of the shear component of the Reynolds stress, $-\overline{u_2'u_1'}(x_2)$, on the mean velocity gradient at all locations, $\partial \overline{U}_1/\partial x_2|_{y_2}$. 
To compute the eddy viscosity, \citet{park2021channel} use the IMFM forcing to maintain the mean velocity gradient, $\partial \overline{V}_1/\partial x_2$, as a Dirac delta function.
In discretized form, $\mathbf{b} = \mathbf{A}{\mathbf{v}}$, where $\mathbf{b} = -\overline{u_2'v_1'}$ is a $N \times 1$ vector, $\mathbf{A} = D_{2121}$ is a $N \times N$ matrix, $\mathbf{v} = \partial \overline{V}_1/\partial x_2$ is a $N \times 1$ vector, and $N$ is the number of degrees of freedom in the averaged space (number of mesh points in $x_2$). 
Using IMFM to specify the velocity gradient as $\mathbf{v} = [1 \ 0 \dots 0]^\top$ (a discrete Dirac delta function) and post-processing the resulting $-\overline{u_2'v_1'}$ from a  simulation of the Navier--Stokes equations (\ref{eq:NS momentum},  \ref{eq:NS mass}) and GMT equations (\ref{eq:GMT momentum}, \ref{eq:GMT mass}) leads to the first column of $\mathbf{A}$. 
Specifying $\mathbf{v} = [0 \ 1 \dots 0]^\top$ leads to the second column, and so forth. 
Thus, obtaining the generalized eddy viscosity using IMFM, or equivalently Hamba's approach~\cite{hamba2005nonlocal} here, requires as many simulations as degrees of freedom in the averaged space. 
In the case of \citet{park2021channel}, 144 simulations were required to produce the eddy viscosity shown in \cref{fig:kernel_D2121}. 
Each simulation solves both incompressible Navier--Stokes and GMT equations, and hence the total cost is equivalent to 288 DNSs. 

\begin{figure}[t]
    \centering
    \includegraphics[scale=1]{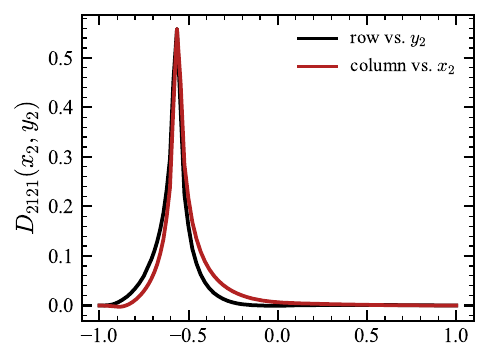}
    \caption{A row of $D_{2121}$ at $x_2=-0.565$, i.e., $D_{2121}(x_2=-0.565,y_2)$ and the corresponding column, $D_{2121}(x_2,y_2=-0.565)$, which shows that $D_{2121}(x_2,y_2)$ is not symmetric (data from \citet{park2021channel} with author permission).
    }
    \label{fig:D2121_row_vs_column} 
\end{figure}

Each simulation obtains a column of $D_{ijkl}$, but the rows of $D_{ijkl}$ are more useful from a physical perspective. 
The rows give the dependence of the Reynolds stress, $-\overline{u_i'u_j'}$, at a given location, on the mean velocity gradient at all locations. 
The rows give information about the importance of nonlocality and regions of mean velocity gradient sensitivity. 
Moreover, the rows of $D_{ijkl}$ are generally not identical to the columns. 
For example, from \cref{fig:kernel_D2121} it may seem that $D_{2121}(x_2, y_2)$ is symmetric and equal to $D_{2121}(y_2, x_2)$. 
\Cref{fig:D2121_row_vs_column} shows clear differences between a row of $D_{2121}$ at $x_2=-0.565$ and a column at the same location. 
We address the need for a method for obtaining targeted rows of the generalized eddy viscosity without first performing a brute-force computation of all columns of the eddy viscosity. 

\section{Adjoint MFM}\label{adjoint MFM}

We develop a method for obtaining a specific row of the generalized eddy viscosity, representing the nonlocal dependence of the Reynolds stress at a specific point on mean velocity gradients at all points in space/time, using an adjoint formulation of the GMT equations. 

The generalized eddy viscosity is part of a linear operator, $\overline{\mathcal{L}}$, that acts on the mean variables, $\overline{V}$, such that the mean equation is
\begin{gather}
    \overline{\mathcal{L}} \, \overline{V}=0.
\end{gather}
For example, $\overline{V}$ is a vector of all mean velocity components and pressure, $\overline{V} = [\overline{V_j} \, \overline{Q}]^\top$, and $\overline{\mathcal{L}}$ includes closed operators and the Reynolds stress closure operator formed by the eddy viscosity, written in block form as
{\renewcommand{\arraystretch}{2}
\begin{gather}
    \overline{\mathcal{L}} = 
    \begin{bmatrix*}[c]
        \dfrac{\partial}{\partial t} + \overline{U}_i\dfrac{\partial}{\partial x_i} - \dfrac{1}{\mathrm{Re}}\dfrac{\partial^2}{\partial x_i \partial x_i} - \dfrac{\partial}{\partial x_i}D_{ijkl}\dfrac{\partial}{\partial x_k} & & \dfrac{\partial}{\partial x_j} \\
        \dfrac{\partial}{\partial x_j} & &  0
    \end{bmatrix*}.
\end{gather}
}
Similarly, the governing equations, e.g, the GMT equations \eqref{eq:GMT momentum} and \eqref{eq:GMT mass}, can be written as
\begin{gather}
    \mathcal{L}v=0.
\end{gather}
\citet{mani2021macroscopic} show that
\begin{gather}
    \label{eq:lin alg MFM relation}
    \overline{\mathcal{L}}=(P\mathcal{L}^{-1}E)^{-1},
\end{gather}
where $P$ is a projection operator such that $\overline{V} = Pv$ and $E$ is an extension operator such that $E = nP^\top$, where $n$ is the number of points used for averaging. 
The derivation of \eqref{eq:lin alg MFM relation} is shown in \cref{lin alg derivation}.
In practice, most problems have a large number of degrees of freedom, and $\mathcal{L}$ is expensive to invert directly, so \citet{park2021channel} use IMFM as described in \cref{problem formulation} to compute the generalized eddy viscosity. 
However, we show \eqref{eq:lin alg MFM relation} to illustrate the relationship between the generalized eddy viscosity embedded in the averaged operator, $\overline{\mathcal{L}}$, and the governing equations. 

The desired rows of the eddy viscosity, $D_{ijkl}$, are the same as the columns of its transpose, $D_{ijkl}^\top$. The transpose $D_{ijkl}^\top$ is linearly embedded in the transpose of the averaged operator,  $\overline{\mathcal{L}}^\top$, which can be computed from \eqref{eq:lin alg MFM relation} according to:
\begin{gather}
    \label{eq:transpose lin alg MFM relation}
    \overline{\mathcal{L}}^\top=(E^\top\mathcal{L}^{-\top}P^\top)^{-1} = (P\mathcal{L}^{-\top}E)^{-1}.
\end{gather}
Equation \eqref{eq:transpose lin alg MFM relation} is similar to \eqref{eq:lin alg MFM relation}, and rather than inverting $\mathcal{L}^\top$ directly, IMFM can be used on $\mathcal{L}^\top$ to compute columns of $D_{ijkl}^\top$. 
This is equivalent to using IMFM on the adjoint of the governing equations. 
The adjoint of the GMT equations in \eqref{eq:GMT momentum} and \eqref{eq:GMT mass} is
\begin{subequations}
    \begin{align}
        \label{eq:adjoint GMT momentum pre time relabeling}
        -\frac{\partial v_i^\dag}{\partial t} - \frac{\partial u_jv_i^\dag}{\partial x_j} &= \frac{\partial q^\dag}{\partial x_i} + \frac{1}{\Rey}\frac{\partial^2 v_i^\dag}{\partial x_j \partial x_j} + s_i, \\
        \frac{\partial v_i^\dag}{\partial x_i} &= 0,
    \end{align}
\end{subequations}
where $v_i^\dag$ and $q^\dag$ are the adjoint velocity and pressure, respectively.
We define a reverse time $\tau \equiv t_f - t$, where $t_f$ is the final simulation time, so 
\begin{subequations}
    \label{eq:adjoint GMT}
    \begin{align}
        \label{eq:adjoint GMT momentum}
        \frac{\partial v_i^\dag}{\partial \tau} - \frac{\partial u_jv_i^\dag}{\partial x_j} &= \frac{\partial q^\dag}{\partial x_i} + \frac{1}{\Rey}\frac{\partial^2 v_i^\dag}{\partial x_j \partial x_j} + s_i, \\
        \label{eq:adjoint GMT mass}
        \frac{\partial v_i^\dag}{\partial x_i} &= 0.
    \end{align}
\end{subequations}
The advective velocity fields, $u_j$, are first computed using a DNS of the incompressible Navier--Stokes equations in forward time order and then read in reverse time order for solving \eqref{eq:adjoint GMT momentum}. It is important to remark that the adjoint equations \eqref{eq:adjoint GMT} are the dual of the GMT system and not the Navier--Stokes equations. As a result, these equations differ from the adjoint Navier--Stokes operator commonly adopted in nonlinear optimization and data assimilation \citep[e.g.][]{wang2019discrete,wang2021state,zaki2021prf}.  Specifically, the difference arises due to the treatment of the advection term, which is linear in the GMT system and is linearized when deriving the adjoint to the Navier--Stokes equations. In this regard, the adjoint GMT equation \eqref{eq:adjoint GMT momentum} is therefore more akin to the adjoint to the scalar transport equation \citep{Wang_hasegawa_zaki_2019}, but also additionally includes the adjoint pressure $q^\dag$ and is accompanied by the divergence-free condition \eqref{eq:adjoint GMT mass}.

A specific row of $D_{ijkl}$ can now be obtained by using IMFM on the adjoint GMT equations in \eqref{eq:adjoint GMT momentum} and \eqref{eq:adjoint GMT mass}. 
In considering the transpose of $D_{ijkl}$, the tensorial components are also transposed such that $D_{ijkl} \rightarrow D_{klij}$. 
For example, consider the discretization of the generalized eddy viscosity of \eqref{eq:generalized eddy viscosity}:
\begin{gather}
\begin{bmatrix}
-\overline{u_1'v_1'} \\
-\overline{u_1'v_2'} \\ 
-\overline{u_1'v_3'} \\ 
-\overline{u_2'v_1'} \\ 
\vdots \\
-\overline{u_3'v_3'}
\end{bmatrix}
=
\begin{bmatrix}
D_{1111} & D_{1112} & D_{1113} & D_{1121} & \dots & D_{1133} \\
D_{1211} & D_{1212} & D_{1213} & D_{1221} & \dots & D_{1233} \\
D_{1311} & D_{1312} & D_{1313} & D_{1321} & \dots & D_{1333} \\
D_{2111} & D_{2112} & D_{2113} & D_{2121} & \dots & D_{2133} \\
\vdots & \vdots & \vdots & \vdots & \ddots & \vdots \\
D_{3311} & D_{3312} & D_{3313} & D_{3321} & \dots & D_{3333} \\
\end{bmatrix}
\begin{bmatrix}
\partial \overline{V}_1/\partial x_1\\
\partial \overline{V}_2/\partial x_1\\
\partial \overline{V}_3/\partial x_1\\
\partial \overline{V}_1/\partial x_2\\
\vdots \\
\partial \overline{V}_3/\partial x_3\\
\end{bmatrix},
\end{gather}
where for each $i,j,k,l \in \{1,2,3\}$, $-\overline{u_i' v_j'}$ is a $N \times 1$ vector, $D_{ijkl}$ is a $N \times N$ block matrix, and $\partial \overline{V_l}/\partial x_k$ is a $N \times 1$ vector where $N$ is the number of degrees of freedom in the averaged space. 
In IMFM, as used by \citet{park2021channel}, forcing one element of $\partial \overline{V}_1/\partial x_2$ to be nonzero and post-processing $-\overline{u_i' v_j'}$ leads to one column in each $D_{ij21}$ matrix. 
Adjoint MFM obtains a row in each $D_{21ij}$ matrix.
If, for example, a row of $D_{1121}$ is desired instead, then one should force an element of $\partial \overline{V}_1^\dag/\partial x_1$ and post-process $\overline{u_2' v_1^{\dag\prime}}$.

\section{Channel setup and numerical details}\label{numerical details}

For DNS of the turbulent channel flow, we use the three-dimensional incompressible Navier--Stokes solver developed by \citet{bose2010grid} and modified by \citet{seo2015pressure}. 
The flow is driven by a nondimensionalized mean pressure gradient, $r_i = (1, 0, 0)$. The Reynolds number, $\Rey_\tau = u_\tau \delta/\nu$, is defined based on the channel half-height, $\delta = 1$, and friction velocity, $u_\tau = 1$.

\citet{park2021channel} modified the solver to include the GMT equations in \eqref{eq:GMT momentum} and \eqref{eq:GMT mass}. 
We modified the solver for the adjoint GMT equations in \eqref{eq:adjoint GMT momentum} and \eqref{eq:adjoint GMT mass}. 
We first conduct a DNS with output fields $u_j$ at each timestep. 
We then solve \eqref{eq:adjoint GMT momentum} by stepping backward in time and reading the $u_j$ fields in reverse order. 
The solenoidal condition in \eqref{eq:adjoint GMT mass} is enforced using a fractional-step method. 
The post-processing involves averaged statistics, not instantaneous flow fields, so the differences between continuous and discrete adjoint formulations are unimportant~\cite{wang_wang_zaki_2022}.

Periodic boundary conditions are enforced in the streamwise ($x_1$) and spanwise ($x_3$) directions, and no-slip and no-penetration boundary conditions are enforced at the walls. 
All solvers use semi-implicit time advancement~\citep{kim1985application}; second-order Crank--Nicolson is used for the wall-normal diffusion terms and Adams--Bashforth is used for all other terms. 
For spatial discretization, the solvers use second-order finite differences on a staggered mesh~\cite{morinishi1998fully} with uniform spacing in $x_1$ and $x_3$ and nonuniform spacing in $x_2$. 
The domain size is $L_1 \times L_2 \times L_3 = 2\pi \times 2 \times \pi$ with $N = 144$ grid cells in each direction. 
The pressure Poisson equation is solved using Fourier transforms in the periodic $x_1$ and $x_3$ directions and a tridiagonal solver in the $x_2$ direction.

\subsection{Obtaining rows of $D_{2121}$}

\citet{park2021channel} specify the mean streamwise velocity as Heaviside functions, $\overline{V}_1 = \theta(x_2-x_2^*)$, at wall-normal locations, $x_2^*$, which are maintained by the forcing. 
This specifies the mean velocity gradient as a Dirac delta function, $\partial \overline{V}_1/\partial x_2 = \delta(x_2-x_2^*)$. 
\citet{park2021channel} then post-processed $-\overline{u_2'v_1'}$ to obtain the column of $D_{2121}$ at $x_2^*$ and repeated the procedure for all $x_2^*$.
 
Using the adjoint formulation, the adjoint mean streamwise velocity is specified as a Heaviside function, and post-processing of $-\overline{u_2'v_1^{\dag\prime}}$ leads to a row of $-D_{2121}$. 
A negative sign accounts for the transposition of $\partial/\partial x_1$ in the velocity gradient.

\subsection{Obtaining rows of other components of $D_{ij21}$}\label{Dij21 details}

For other components, which require maintaining adjoint mean velocity gradient directions other than $\partial \overline{V}_1^\dag/\partial x_2$ as Dirac delta functions, specifying the adjoint mean velocity fields as Heaviside functions may not be mathematically well-posed.
For example, obtaining a row of $D_{1121}$ requires specifying $\partial \overline{V}_1^\dag/\partial x_1$ as a Dirac delta function, $\partial \overline{V}_1^\dag/\partial x_1 = \delta (x_2-x_2^*)$, and post-processing $\overline{u_2' v_1^{\dag\prime}}$. 
An adjoint mean velocity field that satisfies both $\partial \overline{V}_1^\dag/\partial x_1 = \delta (x_2-x_2^*)$ and $\partial \overline{V}_1^\dag/\partial x_2 = 0$ does not exist. 
Thus, we decompose the adjoint velocity field into $v_i^\dag=\overline{V}_i^\dag + v_i^{\dag\prime}$, specify $\partial \overline{V}_1^\dag/\partial x_1$ analytically, and solve the corresponding equation for $v_i^{\dag\prime}$. 
In other words, the decomposition is substituted into the adjoint GMT equation in \eqref{eq:adjoint GMT momentum}:
\begin{gather}
    \frac{\partial \overline{V}_i^\dag}{\partial \tau} + \frac{\partial v_i^{\dag\prime}}{\partial \tau}  - u_j\frac{\partial\overline{V}_i^\dag}{\partial x_j} - u_j\frac{\partial v_i^{\dag\prime}}{\partial x_j} = \frac{\partial q^\dag}{\partial x_i} + \frac{1}{\Rey}\frac{\partial^2 \overline{V}_i^\dag}{\partial x_j \partial x_j} + \frac{1}{\Rey}\frac{\partial^2 v_i^{\dag\prime}}{\partial x_j \partial x_j}+ s_i,
\end{gather}
and $\partial \overline{V}_i^\dag/\partial x_j$ is analytically specified. The IMFM forcing, $s_i$, now maintains $\overline{v_i^{\dag\prime}} = 0.$
For further simplification, the mean temporal term and mean diffusion term may be absorbed by the forcing since they adhere to the property $s_i=\overline{s}_i$:
\begin{gather}
    \label{eq:decomp adjoint GMT momentum}
    \frac{\partial v_i^{\dag\prime}}{\partial \tau}  - u_j\frac{\partial\overline{V}_i^\dag}{\partial x_j} - u_j\frac{\partial v_i^{\dag\prime}}{\partial x_j} = \frac{\partial q^\dag}{\partial x_i} + \frac{1}{\Rey}\frac{\partial^2 v_i^{\dag\prime}}{\partial x_j \partial x_j}+ s_i.
\end{gather}

Continuing the above example for obtaining a row of $D_{1121}$, substituting $\partial \overline{V}_1^\dag/\partial x_1 = \delta(x_2 - x_2^*)$ into \eqref{eq:decomp adjoint GMT momentum} leads to the following equation for $v_i^{\dag\prime}$:
\begin{gather}
    \frac{\partial v_i^{\dag\prime}}{\partial \tau} - u_1\delta(x_2-x_2^*)\delta_{i1} - \frac{\partial u_jv_i^{\dag\prime}}{\partial x_j} = \frac{\partial q^\dag}{\partial x_i} + \frac{1}{\Rey}\frac{\partial^2 v_i^{\dag\prime}}{\partial x_j \partial x_j} + s_i,
\end{gather}
where $\delta_{i1}$ is the Kronecker delta and $s_i$ maintains $\overline{v_i^{\dag\prime}} = 0$. 
We enforce a solenoidal $v_i^{\dag\prime}$ as
\begin{gather}
    \frac{\partial v_i^{\dag\prime}}{\partial x_i} = 0.
\end{gather}

While the solenoidal condition is enforced on $v_i^{\dag\prime}$, the analytically specified adjoint mean velocity gradient may not be solenoidal, e.g., when $\partial\overline{V}_1^\dag/\partial x_1 = \delta(x_2 - x_2^*)$ and all other adjoint mean velocity gradient components are zero.
Because the GMT equations in \eqref{eq:GMT momentum} and \eqref{eq:GMT mass} are linear and ultimately only the superposition of the components of $D_{ijkl}$ is needed for the Reynolds stress tensor, we relax the solenoidal constraint on the adjoint mean velocity gradient to ease computation of the individual components of $D_{ijkl}$ by activating various components of the adjoint mean velocity gradient independently. 
Alternatively, the adjoint mean velocity gradient can be considered an IMFM forcing to the governing equation for $v_i^\dag$ that satisfies the requisite property ($s = \overline{s}$).

\section{Results}\label{results}

\subsection{Eddy viscosity comparison}\label{eddy viscosity comparison}

As an illustrative example, we compare the eddy viscosity at one location, $x_2=-0.565$, obtained using the adjoint formulation with that of \citet{park2021channel} obtained using a brute force approach. 
We chose $x_2=-0.565$, corresponding to row $50$ out of $144$, due to its significant asymmetry in the row versus column as shown in \cref{fig:D2121_row_vs_column}, although we expect the results to hold for all locations.
\citet{park2021channel} averaged over 500 eddy turnover time ($\delta/u_\tau$) for their modeling purposes, whereas we averaged over 115 eddy turnover time, which we found sufficient for verification purposes. 

\begin{figure}[t]
    \centering
    \includegraphics[scale=1]{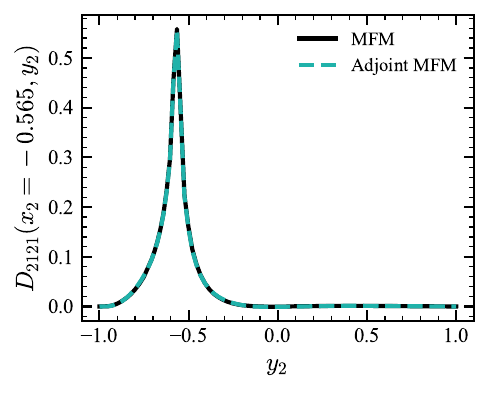}
    \caption{Comparison of $D_{2121}(x_2=-0.565,y_2)$ (corresponding to row $50$ of $144$) using adjoint MFM and from a brute-force calculation by \citet{park2021channel} using MFM.}
    \label{fig:D2121_row} 
\end{figure}

\Cref{fig:D2121_row} shows the eddy viscosity from the adjoint formulation closely matching that of \citet{park2021channel}. 
The normalized error is less than $1\%$. 
We attribute this error to statistical convergence and the shorter averaging times used. 
For example, the normalized error of a regular MFM calculation averaged over $115$ eddy turnover time and compared with the corresponding column is $0.7\%$, and the normalized error of the adjoint MFM calculation used here is $0.8\%$. Therefore, the differences in \cref{fig:D2121_row} are within the uncertainty bounds of the calculation.

\begin{figure}[t]
     \centering
     \begin{subfigure}[t]{0.49\textwidth}
         \centering
         \includegraphics[scale=1]{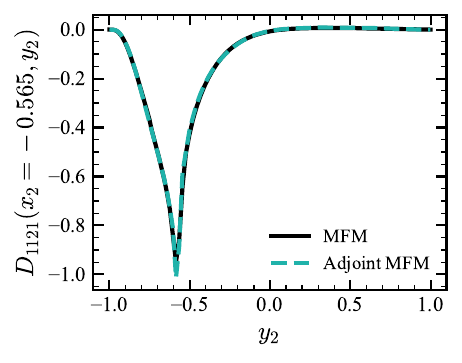}
        \label{fig:D1121_row}
     \end{subfigure}
     \begin{subfigure}[t]{0.49\textwidth}
         \centering
         \includegraphics[scale=1]{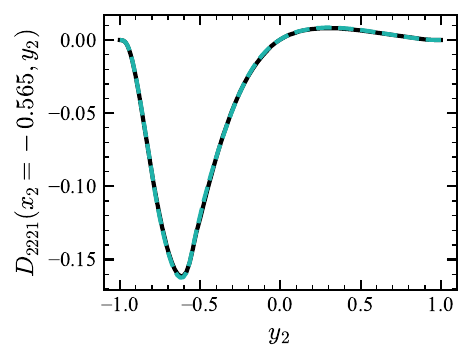}
         \label{fig:D2221_row}
     \end{subfigure}
          \begin{subfigure}[t]{0.49\textwidth}
         \centering
         \includegraphics[scale=1]{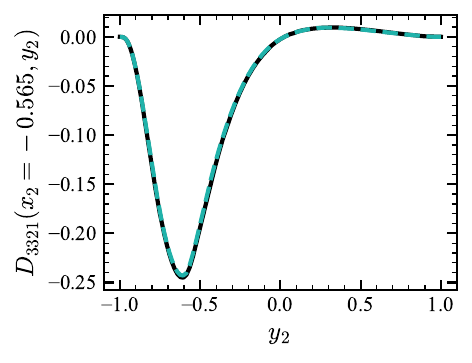}
        \label{fig:D3321_row}
     \end{subfigure}
     \begin{subfigure}[t]{0.49\textwidth}
         \centering
         \includegraphics[scale=1]{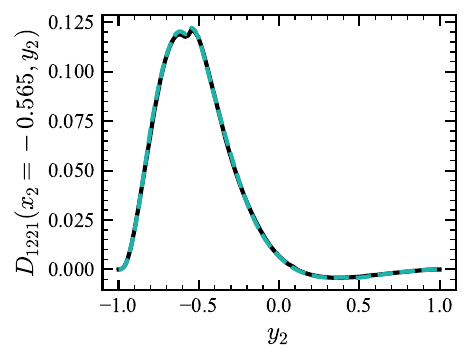}
         \label{fig:D1221_row}
     \end{subfigure}
     \caption{
        Comparison of other components of $D_{ij21}$ at $x_2=-0.565$ (corresponding to row $50$ of $144$) using adjoint MFM and from a brute-force calculation by \citet{park2021channel} using MFM.
     }
     \label{fig:Dij21_row_other_comp}
\end{figure}

\Cref{fig:Dij21_row_other_comp} shows a comparison for the other components of $D_{ij21}$, which are even more asymmetric with regards to rows versus columns as shown in \cref{row column comparison}. 
Due to differences in enforcement of the mean velocity gradient as detailed in \cref{Dij21 details} and the staggered mesh, there is some additional error due to interpolation. 
However, the eddy viscosity from the adjoint formulation still closely matches that of \citet{park2021channel}. 
The largest errors are in $D_{1121}$ due to interpolation of a sharp peak with $4\%$ normalized error. 
The normalized error for all other cases is less than $1.5\%$.

A local approximation is valid if the width of the eddy viscosity kernel is much smaller than the length scale over which the mean velocity gradient varies.
For a channel flow, the mean velocity gradient varies on the order of the channel half-height, $\delta = 1$. A local approximation would model the eddy viscosity as a Dirac delta function such that the Reynolds stress at a given location depends only on the mean velocity gradient at that same location. \Cref{fig:D2121_row} and \cref{fig:Dij21_row_other_comp} show that a local approximation is inappropriate, particularly for the normal components of the Reynolds stress tensor, e.g., $D_{2221}$ corresponding to $-\overline{u_2'u_2'}$ and $D_{3321}$ corresponding to $-\overline{u_3'u_3'}$, that do not exhibit a sharp peak. In all cases, the width of the computed eddy viscosity is $\mathcal{O}(1)$ or on the order of the variation in the mean velocity gradient. Adjoint MFM enables efficient computation of the nonlocal eddy viscosity to analyze these effects for desired regions of the domain. For further characterization of nonlocality and anisotropy in the eddy viscosity, see \citet{park2021channel}.

\subsection{Cost comparison}

A brute force approach to obtaining the eddy viscosity requires as many simulations as degrees of freedom in the averaged space, which for the turbulent channel flow considered in this work is $N = 144$. 
Each simulation solves both Navier--Stokes and GMT equations for a total of 288 DNSs. 
The proposed adjoint simulation uses one simulation per desired eddy viscosity location, which includes a forward solve of the Navier--Stokes equations and a backward solve of the GMT equations for a total of $2$ DNSs. 
Additional overhead is associated with reading and writing the velocity fields to disk for the adjoint simulation and more storage is needed. 

For problems with many degrees of freedom in the averaged space, obtaining the eddy viscosity using a brute-force approach may be computationally intractable. 
However, using a single simulation, the adjoint-based formulation enables targeted quantification of the eddy viscosity at a specific location. 

\section{Conclusion}

The generalized eddy viscosity at a specific location relates the Reynolds stress at that location to mean velocity gradients at all locations, which can be used to characterize nonlocality and sensitivity to the mean velocity gradient. 
In this work, we developed an adjoint-based MFM to cost-effectively compute the eddy viscosity at a specific location of the Reynolds stress using one simulation. Previous brute force approaches~\cite{mani2021macroscopic,hamba2005nonlocal} forced the mean velocity gradient at each location in the averaged space and computed the Reynolds stress response, requiring a separate simulation for each mean velocity gradient location. Hence, these approaches needed as many simulations as degrees of freedom in the averaged space. 

Adjoint MFM can be used to compute the eddy viscosity in regions of interest in turbulent flows, such as at flow separation or reattachment points, to examine nonlocal effects and inform RANS models of deficiencies in their eddy viscosity approximations. A brute force approach would characterize the eddy viscosity for the entire domain (including regions where RANS models perform adequately) and require many simulations to do so. On the other hand, adjoint MFM can be used for more targeted computation of the eddy viscosity in only regions of interest and fewer simulations. 

For applications where the generalized eddy viscosity for the entire domain is still desired, adjoint MFM can also aid in substantially reducing the number of simulations by forcing the mean velocity gradient at selective points that leverage hidden sparsity in the discretized eddy viscosity operator~\cite{bryngelson2023fast}.

\section*{Acknowledgements}

This collaborative effort emerged from support by ONR grants N00014-20-1-2718 (AM), N00014-22-1-2519 (SHB), N00014-23-1-2545 (FS), and N00014-20-1-2715 (TAZ).
JL acknowledges support from the Boeing Company and the computational resources of Lawrence Livermore National Laboratory.
The authors gratefully acknowledge Danah Park for providing the MFM data of the turbulent channel flow simulations.

\appendix

\section{Derivation of the relationship between $\mathcal{L}$ and $\overline{\mathcal{L}}$}
\label{lin alg derivation}
The derivation for the relation in \eqref{eq:lin alg MFM relation} is reproduced from \citet{mani2021macroscopic} below. The governing equations, such as the GMT equations in \eqref{eq:GMT momentum} and \eqref{eq:GMT mass}, can be written as
\begin{gather}
    \label{eq:forced governing eqn}
    \mathcal{L}v=s
\end{gather}
where $v$ is a vector of velocity and pressure, $\mathcal{L}$ is a linear operator, and $s$ is the MFM forcing. 
Similarly, the averaged equations can be written as
\begin{gather}
    \label{eq:forced avgd eqn}
    \overline{\mathcal{L}} \, \overline{V}=\overline{s}
\end{gather}
where $\overline{V}$ is a vector of mean velocity and mean pressure, $\overline{\mathcal{L}}$ is a linear operator, and $\overline{s}$ is the MFM forcing. The averaged operator, $\overline{\mathcal{L}}$, is unknown, and a relation between $\overline{\mathcal{L}}$ and $\mathcal{L}$ is desired. Let averaging be defined by a projection operator, $P$, such that
\begin{gather}
    \label{eq:projection}
    \overline{V}=Pv.
\end{gather}
While the MFM forcing satisfies the property, $s=\overline{s}$, $s$ and $\overline{s}$ may discretely have different dimensions; thus, let $E$ be an extension operator such that
\begin{gather}
    \label{eq:extension}
    s=E\overline{s}.
\end{gather}
Rearranging \eqref{eq:forced governing eqn} and substituting into \eqref{eq:projection} leads to
\begin{gather}
    \overline{V}=P\mathcal{L}^{-1}s=P\mathcal{L}^{-1}E\overline{s},
\end{gather}
where the definition of the extension operator in \eqref{eq:extension} is used. 
Further rearrangement,
\begin{gather}
    (P\mathcal{L}^{-1}E)^{-1} \overline{V} = \overline{s},
\end{gather}
and comparison with \eqref{eq:forced avgd eqn} leads to the relation for $\overline{\mathcal{L}}$ in \eqref{eq:lin alg MFM relation}.

\section{Row versus column comparison for $D_{ij21}$}
\label{row column comparison}
\begin{figure}[H]
     \centering
     \begin{subfigure}[t]{0.49\textwidth}
         \centering
         \includegraphics[scale=1]{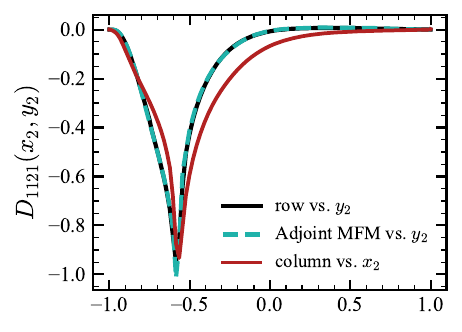}
     \end{subfigure}
     \begin{subfigure}[t]{0.49\textwidth}
         \centering
         \includegraphics[scale=1]{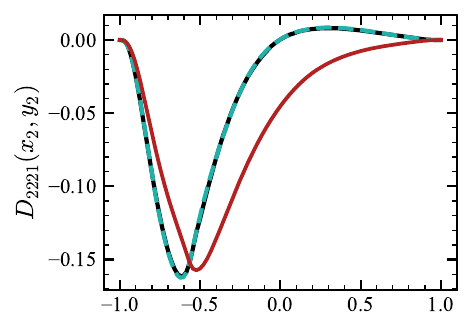}
     \end{subfigure}
          \begin{subfigure}[t]{0.49\textwidth}
         \centering
         \includegraphics[scale=1]{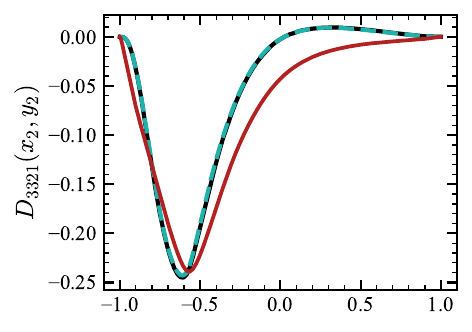}
     \end{subfigure}
     \begin{subfigure}[t]{0.49\textwidth}
         \centering
         \includegraphics[scale=1]{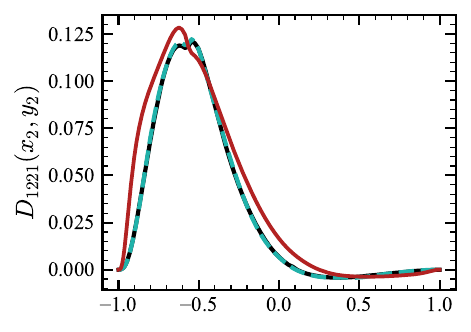}
     \end{subfigure}
     \caption{
        Comparison of row versus column for $D_{ij21}$ components using data from \citet{park2021channel}. The corresponding row computed using adjoint MFM is also shown.
     }
     \label{fig:Dij21_row_other_comp_w_column}
\end{figure}

In general, the rows of the eddy viscosity are not identical to the columns. \Cref{fig:D2121_row_vs_column} showed the asymmetry in row versus column for $D_{2121}$ at $x_2=-0.565$. \Cref{fig:Dij21_row_other_comp_w_column} shows the asymmetry in row versus column for other components of $D_{ij21}$ at $x_2=-0.565$. Adjoint MFM results from \cref{fig:Dij21_row_other_comp} are also plotted for comparison.

\bibliographystyle{bibsty}
\bibliography{bibfile}

\end{document}